\documentstyle[prl,aps,epsfig]{revtex}
\draft
\begin{document}
\twocolumn[\hsize\textwidth\columnwidth\hsize\csname@twocolumnfalse\endcsname
\title{X-Ray Resonant Scattering as a Direct Probe of Orbital Ordering
in Transition-Metal Oxides}
\author{Michele Fabrizio$^{(a)}$, Massimo Altarelli$^{(b)}$,
and Maurizio Benfatto$^{(b,c)}$}
\address{$^{(a)}$ International School for Advanced Studies (SISSA) 
and
INFM, Via Beirut 2-4, 34013 Trieste, Italy}
\address{$^{(b)}$ European Synchrotron Radiation Facility,
Bo\^{i}te Postale 220, F-38043 Grenoble C\'edex, France}
\address{$^{(c)}$ INFN -- Laboratori Nazionali di Frascati CP 13, 
00044 Frascati, Italy}
\date{September 1997}
\maketitle 
\begin{abstract}
X-ray resonant scattering at the
K-edge of transition metal oxides is shown to
measure the orbital order parameter, supposed to accompany
magnetic ordering in some cases. Virtual transitions to the
$3d$-orbitals are quadrupolar in general.
In cases with no inversion symmetry,  such as V$_2$O$_3$, treated
in detail here, a dipole component enhances the resonance.
Hence, we argue that the detailed structure of orbital order in
V$_2$O$_3$ is experimentally accessible.
\end{abstract}
\pacs{PACS number: 78.70.Ck, 71.30.+h, 71.20.Be}
\vspace{0.2in}
]
\narrowtext
The orbital ordering in compounds containing ions with unfilled shells
is a long standing problem\cite{Khomskii}
and has recently attracted new attention given the interest in the
physics of transition metal oxides stimulated by cuprate 
superconductors and
colossal magnetoresistance materials.
In fact, while many experiments give direct information on magnetic
structures, which are also consequence of partially empty shells, only
indirect evidence points to the existence and structure of orbital 
order,
unless it
is accompanied by a cooperative Jahn-Teller effect with a lattice 
distortion.
An important and intriguing example
is the prototype Mott-Hubbard insulator V$_2$O$_3$\cite{review}.
At ambient pressure,
this compound undergoes a metal-insulator transition at
$T_c\simeq 150K$\cite{McWhan&Remeika}.
The insulating phase below $T_c$ is antiferromagnetically ordered with
a quite peculiar structure having both ferro- and antiferromagnetic
bonds\cite{Moon}. Castellani, Natoli and Ranninger\cite{CN&R}
argued that such a magnetic ordering implies an orbital ordering.
Quite recently, Bao {\it et al.}\cite{Bao} have found by
neutron diffraction that spin correlations in the
metallic phase behave quite differently from the insulating one,
giving further evidence of the importance of the orbital degrees of 
freedom.

Actually,
Castellani {\it et al.}\cite{CN&R} showed that (at least) two
orbital structures, with different order parameters and wavevectors,
are compatible with the observed magnetic one.
Later, based on the exchange constants fitted to neutron scattering
data, one of the two orbital structures came to be
favored\cite{Word,review,Bao}.
However, there is still no direct observation of the orbital ordering,
nor a firm experimental determination of its wavevector, and of the
temperature at which it appears, which may not coincide
with the N\'eel temperature.
In addition to V$_2$O$_3$, there are many other examples where,
in spite of the absence of direct evidences, the orbital degrees
of freedom are believed to play an important role, as the quite 
popular
colossal magnetoresistance material La$_{1-x}$(Ca,Sr)$_x$MnO$_3$.

In this Letter,
we argue that it is indeed possible to
have direct experimental access to the orbital ordering
by X-ray resonant scattering, and that this process
should be sufficiently intense to be readily observable with modern
synchrotron radiation sources.

First of all we notice that, contrary to what we are going to argue
for the resonant scattering,  the non resonant X-ray diffraction has 
extremely
little sensitivity to orbital order. We have estimated the cross 
section
for orbital order scattering in this case
by using hydrogenic $3d$-wavefunctions, and we find the corresponding
scattering power at the orbital order wavevector to be less then 0.1
electrons per unit cell.

Let us now show how an X-ray resonant elastic scattering
experiment can detect orbital ordering in transition metal compounds.
The energy of the incoming beam is supposed to resonate with the
K-edge of the transition element.
Virtual transitions to the conduction band $d$-like components are
quadrupolar, and the corresponding transition operator at a given atom
in the lattice takes the form:
\begin{eqnarray}
&-& C_2 \sqrt{n_{\epsilon,k}} \frac{c}{\omega} \sum_\sigma \left[
\sqrt{3}\,\epsilon_zk_z\,d^\dagger_{3z^2\!-\!r^2\!,\sigma}
\!\!+\!
(\epsilon_xk_x\!-\!\epsilon_yk_y)d^\dagger_{x^2\!-\!y^2\!,\sigma} 
\right.
\nonumber\\
&+&\! (\epsilon_xk_y\!+\!\epsilon_yk_x)d^\dagger_{xy,\sigma}
\!+\!  (\epsilon_xk_z\!+\!\epsilon_zk_x)d^\dagger_{xz,\sigma}
\nonumber\\
&+&\left.\! (\epsilon_yk_z\!+\!\epsilon_zk_y)d^\dagger_{yz,\sigma}
\right] s_\sigma
\equiv - C_2 \sqrt{n_{\epsilon,k}}
\sum_\sigma \hat{D}^\dagger_{\sigma}(\epsilon,k) s_\sigma,
\label{F2}
\end{eqnarray}
which also defines the dimensionless absorption operator
$\hat{F} \equiv \sum_\sigma \hat{D}^\dagger_{\sigma } s_\sigma$.
In the above equation,  $d^\dagger_{i,\sigma}$ is the operator which
creates an electron of spin $\sigma$ in the
$i$ $d$-orbital,
while $s_\sigma$ creates a hole in the $1s$-shell, and
$C_2 = (e \omega^2 r_{2,ds}/c) \sqrt{2\pi\hbar/(15 \omega)}$,
with $\omega$ the photon
frequency, and $n_{\epsilon,k}$ the
density in the incoming beam of photons with polarization
$\vec{\epsilon}$
and wavevector $\vec{k}$. The radial matrix element is defined by
$r_{2,ds} = \int r^2dr \chi_d^*(r) r^2 \chi_s(r)$,
in terms of the radial wavefunctions $\chi_s$ and $\chi_d$.

If the atom is at position $\vec{R}$, a phase factor
$
{\rm e}^{i\vec{k}\cdot \vec{R}}
$
weights the contribution of this single process.
In obvious notation,
the absorption operator at this site is denoted by $\hat{F}_{R}$.
The transition probability amplitude of an elastic scattering in which the
above absorption process is followed by the emission of a photon
with the same energy but polarization $\vec{\epsilon'}$ and
wavevector $\vec{k'}$ is given by
\begin{equation}
C_2^2 \sqrt{\frac{n_{\epsilon,k}}{V}}\!\!\! \sum_{n,\vec{R},\vec{R'}}
\!\!\!\!{\rm e}^{i(\vec{k}\cdot\vec{R}-\vec{k'}\cdot\vec{R'})}
\frac{ \langle g | \hat{F}_{R'}^\dagger(\epsilon',k') | n\rangle
\langle n | \hat{F}_{R} (\epsilon,k) | g \rangle }
{\hbar\omega  - \hbar\omega_{ng} + i\Gamma},
\label{AP1}
\end{equation}
where $|g\rangle$ is the ground state, $|n\rangle$ an intermediate 
state
and $\Gamma/\hbar$ the inverse lifetime of the core-hole due to
other decay processes which might occur before the emission process
takes place.

At resonance
(\ref{AP1}) becomes through (\ref{F2})
\begin{equation}
-i\frac{C_2^2}{\Gamma}\sqrt{\frac{n_{\epsilon,k}}{V}}
\sum_{\sigma, \vec{R}} {\rm e}^{i\vec{Q}\cdot\vec{R}}
\langle g |\hat{D}_{\sigma, R}(\epsilon',k')
\hat{D}^\dagger_{\sigma, R}(\epsilon,k)
| g\rangle.
\label{def-O}
\end{equation}
In reality the resonant energy is not sharply defined, but the width
of the $d$-like bands involved in the ordering is comparable  to 
typical
monochromator resolutions defining the incoming energy and to the
lifetime broadening  $\Gamma\simeq 1eV$.
Therefore the resonant scattering via a K-edge quadrupole transition
is a measure of the Fourier
component at momentum $\vec{Q}=\vec{k}-\vec{k'}$ of an operator
\begin{equation}
\hat{O} = \sum_\sigma \hat{D}_{\sigma}(\epsilon',k')
\hat{D}^\dagger_{\sigma}(\epsilon,k),
\label{O}
\end{equation}
which is a combination of products of
d-orbitals annihilation and creation operators. The scattering
geometry defines uniquely such a combination.
The form of the scattering operator (\ref{O}) is similar to
that of the orbital order parameter, generally written as
$\hat{\Delta} = \sum_{i,j} a_{ij} d^\dagger_i d_j$,
where $a_{ij}$ depend on the system under consideration (see below the
specific case of V$_2$O$_3$). For a suitable choice of the scattering 
geometry,
 i.e. of wavevectors and polarizations, the scattering operator
(\ref{O}) includes a component equal to the orbital order parameter.
This component will result in the only non vanishing amplitude for
momentum transferred $\vec{Q}$ equal to the orbital order wavevector.
The treatment sketched here can be applied without modification to
cases like LaMnO$_3$, in which the quadrupolar scattering channel is 
the
only available one. In the case of V$_2$O$_3$, as we shall see, the 
absence
of inversion symmetry at V-sites in the insulating phase allows dipole
transitions as well. We will consider this more complicated but
interesting case in detail.

According to the general wisdom, the V ions in this compound have an
oxidation state V$^{3+}$, thus containing two 
$d$-electrons\cite{review}.
The five $d$-orbitals are split into a lower triplet of t$_{2g}$ 
orbitals
and an higher doublet e$_g$. It is common to use a reference frame
in which the $z$-axis coincides with the $c$-axis of the non primitive
hexagonal cell, while the $x$-axis is parallel to one of the bond
connecting the V-atoms in the honeycomb lattice of the $ab$-plane.
In this reference frame, the three $t_{2g}$ orbitals become:
$d_1 = d_{3z^2-r^2}$ ,and
\begin{equation}
d_2 = \sqrt{\frac{2}{3}}d_{xy} + \sqrt{\frac{1}{3}}d_{xz},
d_3 = -\sqrt{\frac{2}{3}}d_{x^2-y^2} - \sqrt{\frac{1}{3}}d_{yz}.
\label{d3}
\end{equation}
The $d_1$ orbital points towards the only nearest neighbor V along the
$c$-axis, thus forming a strong covalent bond which is filled. Hence
the $d_1$ orbital, being inert, does not participate the
orbital ordering, which only involves the $d_2$ and $d_3$ orbitals.
We notice that these orbitals,
because of the crystal field in the corundum structure,
acquire also a component of $p$-like symmetry, which we need to
identify. A V-atom is surrounded by a distorted oxygen octahedron
and by four vanadium second neighbors, providing a non symmetric
environment, as described in Ref.\cite{Marezio}.
The crystal field potential on the central vanadium
due to this environment may be parametrized by
\begin{equation}
V_{cf} = V_1 \left( Y_{3,-3} - Y_{3,3} \right) + V_2 Y_{3,0}.
\label{Vcf}
\end{equation}
As a result, the modified $d$-orbitals involved in the
orbital ordering are, apart from the normalization,
\begin{equation}
d_2 \to d_2 + \eta(u p_y + v p_x)\; ; \; 
d_3 \to d_3 + \eta(u p_x - v p_y), 
\label{dp}
\end{equation}
where $\eta u$ and $\eta v$ ($u^2+v^2=1$) are proportional
to $V_1$ and $V_2$, respectively,
and inversely proportional to the energy difference between the
vanadium $3d$-orbitals and the $p$-orbitals involved.
In order to obtain an estimate of this hybridization, an X$\alpha$
calculation in the $Z+1$ approximation for a VO$_6$ cluster was
performed, and provided $\eta\simeq 0.2$, and $u\simeq -0.37$.
A test of the adequacy of this approximation for a quantitative 
estimate
is given by comparing the energy difference between the lowest 
absorption
peak related to the $d-p$ hybridized levels and the main absorption 
edge,
which is 20.7$eV$ in the cluster calculation, and 20.1$eV$
experimentally\cite{Wong}. Also the relative intensities are in
satisfactory agreement with experiment.

According to Ref.\cite{CN&R}, two orbital orderings are compatible 
with the
observed magnetic structure.
For the first one, the staggered orbital parameter is proportional to
the operator
\begin{equation}
{\rm case~(A):}\;\; \hat{\Delta}_A = d^\dagger_2 d_3 + d^\dagger_3 
d_2.
\label{delta1}
\end{equation}
In the hexagonal reciprocal lattice, the $\vec{Q}$-vector of this
orbital ordering is predicted to be different from that of the 
magnetic
ordering.
In fact, while the latter has maximum diffraction at 
$\vec{Q}=(0.5,0.5,0)$
and zero at $\vec{Q}=(0.5,0.5,1)$, for the former the situation
is reversed. This makes the search for the orbital ordering easier.
In Cartesian coordinates, the
wave vector of the orbital order (A) is
\begin{equation}
\vec{Q}_A = 2\pi \left( 0, \frac{1}{a\sqrt{3}} , \frac{1}{3c}\right),
\label{Q1-vec}
\end{equation}
where $a$ is the bond length in the honeycomb lattice
of the basal plane, and $c$ is the distance between two
nearest neighbor V-atoms in the z-direction. The hexagonal cell
that we use contains three layers, which is compatible with the
notations used by Moon\cite{Moon} and McWhan and 
Remeika\cite{McWhan&Remeika}.

The other orbital order compatible with the magnetic structure has an
order parameter proportional to
\begin{equation}
{\rm case~(B):}\;\; \hat{\Delta}_B = d^\dagger_2d_2 - d^\dagger_3 d_3.
\label{delta2}
\end{equation}
For this case the situation is even simpler, since the
ordering is no more staggered in the basal plane, while it
remains staggered along the $c$-direction. However, since the two 
V-atoms
on the basal plane belonging to the same hexagonal cell have opposite
values of the order parameter, the smallest $\vec{Q}$ at which we 
expect a
diffraction peak is
\begin{equation}
\vec{Q}_B = 2\pi \left(\frac{1}{3a},\frac{1}{a \sqrt{3}},
\frac{1}{3c}\right).
\label{Q2-vec}
\end{equation}

Let us now apply the general treatment described above to this
specific example. However, since the conduction band has both $d$ and
$p$-like character, we have to extend the general formalism to 
include,
besides the quadrupole, also
a dipole component in the transition amplitude.

First of all, let us consider the quadrupole channel already 
discussed.
Since only the $d_2$ and $d_3$ combinations of Eq.(\ref{d3})
are involved in the orbital ordering, it is more useful
to pick up from (\ref{F2}) only the part involving these
orbitals. All the other allowed transitions are not expected to give a
contribution to the elastic scattering at that particular
$\vec{Q}$. Hence, the relevant quadrupole absorption
operator simplify to
\begin{equation}
\hat{F}_2 = \frac{1}{\sqrt{1+\eta^2}} \frac{c}{\omega} \sum_\sigma 
\left[
\left(\vec{\epsilon}\cdot\hat{M}_2\vec{k}\right) d^\dagger_{2\sigma}
- \left(\vec{\epsilon}\cdot \hat{M}_3\vec{k}\right)
d^\dagger_{3 \sigma}\right] s_\sigma ,
\label{abs-op}
\end{equation}
where $1/\sqrt{1+\eta^2}$ weighs the $d$-component in the conduction
band [see Eq.(\ref{dp})], and the matrices $M$ are
\[
\hat{M}_2 =
\left(
\begin{array}{ccc}
0 & \sqrt{2} & 1 \\
\sqrt{2} & 0 & 0 \\
1 & 0 & 0 \\
\end{array}
\right),\,
\hat{M}_3 =
\left(
\begin{array}{ccc}
\sqrt{2} & 0 & 0 \\
0 & -\sqrt{2} & 1 \\
0 & 1 & 0 \\
\end{array}
\right).
\]

Next, let us study the dipole channel.
Using the same notations as above, the dimensionless dipole absorption
operator at resonance
$\hat{F}_1= \sum_\sigma \hat{P}^\dagger_\sigma s_\sigma$ is defined
through the dimensional one by
\begin{equation}
i C_1 \sqrt{n_{\epsilon,k}} \sum_{j=x,y,z}\sum_\sigma
\epsilon_j p^\dagger_{j,\sigma} s_\sigma
\equiv
i C_1 \sqrt{n_{\epsilon,k}} \sum_\sigma \hat{P}^\dagger_\sigma 
s_\sigma,
\label{F1}
\end{equation}
where $p^\dagger_{j,\sigma}$
creates a spin-$\sigma$ electron in the $j$ $p$-orbital, and
$C_1 = e \omega r_{1,ps} \sqrt{2\pi\hbar/(3\omega)}$, where
$r_{1,ps} = \int r^2 dr \chi^*_p(r) r \chi_s(r)$,
in analogy with the definition of $r_{2,ds}$.

The component of the dipole absorption operator $\hat{F}_1$ which is 
sensible
to the orbital ordering can be easily found
through Eq.(\ref{dp}), namely
\begin{equation}
\hat{F}_1 = \frac{\eta}{\sqrt{1+\eta^2}}
\sum_\sigma \left[
\left(\vec{v}_2 \cdot \vec{\epsilon}\right)\, d^\dagger_{2,\sigma}
+\left(\vec{v}_3 \cdot \vec{\epsilon}\right)\,
d^\dagger_{3,\sigma}\right] s_\sigma,
\label{dipole}
\end{equation}
where the factor in front of the sum is the weight of the
$p$-component in the conduction band wave functions, and,
for shortness, we have introduced the two vectors
$\vec{v}_2 = (v,u,0)$ and $\vec{v}_3 = (u,-v,0)$.

The whole transition operator is the sum of the quadrupole and dipole
components. However Eqs.(\ref{AP1}), (\ref{def-O}) and (\ref{O}) are
still correct, provided we take
\begin{equation}
\hat{F} = \frac{1}{\sqrt{1+\eta^2}}
\left[ \hat{F}_2 + i\eta \frac{C_1}{C_2} \hat{F}_1\right].
\label{F-V2O3}
\end{equation}
As a result, the scattering cross section will include, besides
pure quadrupole and dipole contributions, an interference term.
To further proceed, we need to extract from (\ref{O}) the term
proportional to the orbital order parameter, which can be
in general written as
$\left( T^{\alpha}_{11} + T^{\alpha}_{22} -i T^{\alpha}_{12}
\right)\hat{\Delta}_\alpha$.
Here $\alpha=A,B$ refers to the two possible order parameters
Eq.(\ref{delta1}) and Eq.(\ref{delta2}), and $T^\alpha_{11}$,
$T^\alpha_{22}$, $T^\alpha_{12}$ indicate the contributions of
the pure dipole, the pure quadrupole and the interference channels,
respectively. For these polarization and wavevector dependent
quantities, we find:
\begin{eqnarray}
T^A_{22} &=& \frac{c^2}{\omega^2(1+\eta^2)}
\left[
\left(\vec{\epsilon'}^*\hat{M}_2\vec{k'}\right)\!
\left(\vec{\epsilon}\hat{M}_3\vec{k}\right) +
(2\leftrightarrow  3) \right],
\label{TA22} \\
T^A_{11} &=& \frac{\eta^2}{1+\eta^2}\frac{C_1^2}{C_2^2}
\left[
\left(\vec{\epsilon'}^*\!\cdot\vec{v_2}\right)\!
\left(\vec{\epsilon}\cdot\vec{v}_3\right) +
(2\leftrightarrow 3)
\right],
\label{TA11}\\
T^A_{12} &=& \frac{\eta}{1+\eta^2}\frac{cC_1}{2\omega C_2}
\left[
\left(\vec{\epsilon'}^*\hat{M}_2\vec{k'}\right)\!
\left(\vec{\epsilon}\cdot\vec{v}_3\right)\! +\!
\left(\vec{\epsilon'}^*\!\!\cdot\vec{v}_2\right)\!
\left(\vec{\epsilon}\hat{M}_3\vec{k}\right) \right.\nonumber\\
& & \left.
\phantom{\left(\vec{\epsilon'}^*\hat{M}_2\right)}
- (2\leftrightarrow 3)
\right],
\label{TA12}\\
T^B_{22} &=& \frac{c^2}{\omega^2(1+\eta^2)}
\left[
\left(\vec{\epsilon'}^*\hat{M}_2\vec{k'}\right)
\left(\vec{\epsilon}\hat{M}_2\vec{k}\right) -
(2\leftrightarrow 3)\right],
\label{TB22}\\
T^B_{11} &=& \frac{\eta^2}{1+\eta^2}\frac{C_1^2}{C_2^2}
\left[
\left(\vec{\epsilon'}^*\cdot\vec{v_2}\right)
\left(\vec{\epsilon}\cdot\vec{v}_2\right) -
(2\leftrightarrow 3)
\right],
\label{TB11}\\
T^B_{12} &=& \frac{\eta}{1+\eta^2}\frac{c C_1}{2\omega C_2}
\left[
\left(\vec{\epsilon'}^*\hat{M}_2\vec{k'}\right)\!
\left(\vec{\epsilon}\cdot\vec{v}_2\right)\! +\!
\left(\vec{\epsilon'}^*\!\!\cdot\vec{v}_2\right)\!
\left(\vec{\epsilon}\hat{M}_2\vec{k}\right) \right.\nonumber\\
& &\left.
\phantom{\left(\vec{\epsilon'}^*\hat{M}_2\right)}
+ (2\leftrightarrow 3)
\right].
\label{TB12}
\end{eqnarray}
From the cluster calculation, we obtain the radial matrix elements
for dipole and quadrupole transitions, which allow us to establish 
that
the relative weight of the two terms in Eq.(\ref{F-V2O3}) is given by
$\eta C_1/C_2 \simeq 7.4$.
Therefore we expect the pure dipole term alone to reproduce the total
cross-section within $\simeq 13\%$ accuracy. Hence, for the purpose
of the present paper, it will be sufficient to calculate just this 
term.

We are now in position to describe
the behavior of the cross section for orbital order scattering
as a function of experimental configuration.
Since the scattering is elastic and the
transferred momentum is fixed, also the scattering angle
$2 \gamma$ is determined by
$\sin \gamma = Qc/(2\omega)$,
being $\omega$ the photon frequency, and $Q$ the modulus of the
wavevector for the orbital ordering, which will be either
(\ref{Q1-vec}) or (\ref{Q2-vec}).
Taking the K-edge of the vanadium $\hbar\omega = 5465 eV$, and the
lattice parameters in the insulating phase $a=2.88 \AA$ and $c= 2.70 
\AA$,
we find the two scattering angles $\gamma_A=15.49^\circ$ and
$\gamma_B=17.34^\circ$.
If we assume not to detect the polarization of the emitted light,
only three degrees of freedom
for the scattering geometry remain. The first
is the angle $\phi$ of rotation
of the scattering plane around the transferred
momentum $\vec{Q}$, with $\phi=0$ corresponding to the component of 
the
incoming wavevector $\vec{k}$ normal to $\vec{Q}$ lying in the
$ab$-plane. The other two free scattering parameters are related to
the polarization of the incoming beam, and reduce to one for a
linearly polarized light.

Let us now estimate the
scattering cross section. The radial dipole matrix element as obtained
from the cluster calculation is $r_{1,ps}=780 r_0$, where $r_0$ is
the classical electron radius.  We also take
$\Gamma= 0.8eV$\cite{Benfatto}. The resulting cross section is very 
large,
of order $10^3 r_0^2$, which is not surprising in view of our 
assumption
of a single resonating level (similarly large cross sections for 
magnetic
X-ray scatterings were estimated by Hannon {\it et al.}\cite{Hannon},
for the $3d$ to $4f$ resonance of a rare earth with one $f$-hole per
atom). The dependence of the cross section on the angle $\phi$ is
shown in Fig.1 for the orbital order (A) of Eq.(\ref{delta1}) and
in Fig.2 for the case (B) of Eq.(\ref{delta2}).
\begin{figure}
\centerline{\epsfig{file=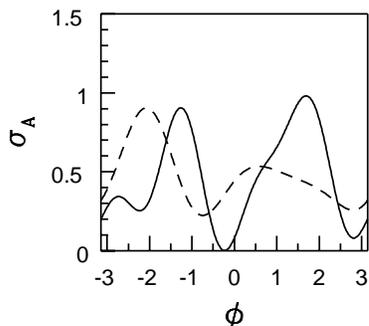,height=4.5truecm}}
\caption{
Orbital order (A) scattering cross section in arbitrary units for 
the $\sigma$-polarization (solid line) and $\pi$-polarization (dashed line).}
\end{figure}

In conclusion it was shown that elastic resonant X-ray scattering in
an appropriate experimental configuration provides a direct
probe of the orbital order parameter. Numerical estimates show that,
in general, the effect should be observable in transition metal oxides
at the metal K-edge. This is even more so in the particularly 
interesting
case of V$_2$O$_3$, where the absence of inversion symmetry in the 
insulating
phase allows dipole transitions to relevant intermediate states. 
Interestingly, the same dipole matrix elements are expected to enhance 
the resonant magnetic scattering from the antiferromagnetic structure, 
allowing to monitor the interplay of orbital and magnetic order in the 
same experiment.
\begin{figure}
\centerline{\epsfig{file=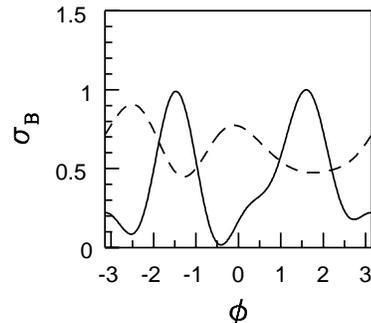,height=4.5truecm}}
\caption{
Orbital order (B) scattering cross section in arbitrary units for 
the $\sigma$-polarization (solid line) and $\pi$-polarization (dashed line).}
\end{figure}

Our results imply that resonant X-ray experiments can determine
the temperature dependence of the orbital order
parameter, as well as its wavevector. These quantities
have so far eluded any direct measurement.

One of us (M.F.) thanks 
ESRF, where most of this work was performed, for its hospitality

\end{document}